**Control of Covalent Bond Enables Efficient Magnetic Cooling**

*Xin Tang*, Yoshio Miura, Noriki Terada, Enda Xiao, Shintaro Kobayashi, Allan Döring, Terumasa Tadano, Andres Martin-Cid, Takuo Ohkochi, Shogo Kawaguchi, Yoshitaka Matsushita, Tadakatsu Ohkubo, Tetsuya Nakamura, Konstantin Skokov, Oliver Gutfleisch, Kazuhiro Hono and Hossein Sepehri-Amin*

Dr. X. Tang, Prof. Y. Miura, Dr. N. Terada, Dr. E. Xiao, Dr. T. Tadano, Dr. A. Martin-Cid, Dr. Y. Matsushita, Prof. T. Nakamura, Dr. T. Ohkubo, Prof. K. Hono, Prof. H. Sepehri-Amin
National Institute for Materials Science
Tsukuba, 305-0047, Japan

Prof. Y. Miura
Faculty of Electrical Engineering and Electronics, Kyoto Institute of Technology,
Matsugasaki, Sakyo-ku, Kyoto, 606-8585, Japan

Dr. S. Kobayashi, Dr. T. Ohkochi and Dr. S. Kawaguchi
Japan Synchrotron Radiation Research Institute, 1-1-1 Kouto, Sayo 679-5198, Japan

Dr. A.M. Döring, Dr. K.P. Skokov, Prof. O. Gutfleisch
Institute of Materials Science, Dep. of Functional Materials, Technical University of Darmstadt Peter-Grünberg Str. 16, Darmstadt, 64287, Germany

Dr. T. Ohkochi
Laboratory of Advanced Science and Technology for Industry, University of Hyogo,
Kamigori, Ako, Hyogo, 678-1205, Japan

Prof. T. Nakamura and Prof. H. Sepehri-Amin
International Center for Synchrotron Radiation Innovation Smart (SRIS), Tohoku University, Sendai 980-8577, Japan

* Corresponding author*: Email address:* Tang.Xin@nims.go.jp (Xin Tang).

Magnetic cooling, harnessing the temperature change in matter when exposed to a magnetic field, presents an energy-efficient and climate-friendly alternative to traditional vapor-compression refrigeration systems, with a significantly lower global warming potential. The advancement of this technology would be accelerated if irreversible losses arising from hysteresis in magnetocaloric materials were minimized. Despite extensive efforts to manipulate crystal lattice constants at the unit-cell level, mitigating hysteresis often compromises cooling performance. Herein, we address this persistent challenge by forming $Sn(Ge)_3$–$Sn(Ge)_3$ bonds within the unit cell of the $Gd_5Ge_4$ compound. Our approach enables an energetically favorable phase transition, leading to the elimination of thermal hysteresis. Consequently, we achieve a synergistic improvement of two key magnetocaloric figures of merit: a larger magnetic entropy change and a twofold increase in the reversible adiabatic temperature change (from 3.8 to 8 K) in the $Gd_5Sn_2Ge_2$ compound. Such synergies can be extended over a wide temperature range of 40–160 K. This study demonstrates a paradigm shift in mastering hysteresis toward simultaneously achieving exceptional magnetocaloric metrics and opens up promising avenues for gas liquefaction applications in the longstanding pursuit of sustainable energy solutions.

## 1. Introduction

The drive toward a carbon-neutral society, prompted by the environmental impact of traditional energy-intensive vapor-compression cooling technologies[1], mandates the exploration of green alternatives with zero carbon emissions. Solid-state caloric refrigeration technologies[2], including electrocaloric[3], elastocaloric[4], and magnetocaloric[5-8] cooling, have thus emerged as promising solutions. Among them, the higher cooling potential of magnetocaloric refrigeration has made it a front-runner to replace conventional vapor-compression refrigeration technology[9]. Utilizing the magnetocaloric effect (MCE), this technology exploits the degree of freedom of magnetic dipoles in materials to induce an isothermal magnetic entropy change ($\Delta S_m$) or adiabatic temperature change ($\triangle T_{ad}$) [6,8]. Despite its potential applications, the commercial viability of magnetocaloric refrigeration is hampered by the scarcity of high-performance magnetic refrigerants.

The discovery of a giant MCE (GMCE) in $Gd_5(Ge,Si)_4$ compound marked a major milestone[10], and triggered extensive search for GMCEs in various intermetallic compounds, such as $(Mn,Fe)_2P$[11], MnAs[12], the Ni-Mn-based Heusler compounds[5], and $La(Fe,Si)_{13}H$[13,14]. These GMCEs arise from the nature of the first-order magnetic phase transitions (FOMTs). In FOMT materials, the magnetic and structural degrees of freedom are strongly coupled, resulting in latent heat and its associated hysteresis. This introduces unwanted irreversibility and mechanical instability, making FOMT materials impractical for applications[8,15,16]. A conventional wisdom to mitigate the hysteresis is to weaken the magnetostructural phase transition by tuning the crystallographic compatibility of two respective phases[17–21]. This is typically achieved by suppressing changes in the symmetry or lattice constants of a unit cell. Despite numerous endeavors, this strategy has been plagued by a persistent tradeoff: sacrificing a large MCE for a small hysteresis. For instance, microalloying Fe in $Gd_5Ge_2Si_2$ compound achieves minimal hysteresis by suppressing structural changes[18], but this reduces $-\Delta S_m$ from 18 to 7 $J{\cdot}kg^{-1}{\cdot}K^{-1}$. Hysteresis can be further reduced by tuning the nature of transition toward second-order magnetic phase transition (SOMT) in Si-rich $Gd_5(Ge_{1-x}Si_x)_4$ compounds[22], wherein ferromagnetic ordering occurs without an alteration in the crystal structure of $Gd_5Si_4$ compound[23], enabling a reversible MCE; however, this reduces the giant value of $-\Delta S_m$ to 6 $J{\cdot}kg^{-1}{\cdot}K^{-1}$ due to elimination of latent heat. This tradeoff is commonly observed across many FOMT compounds[18,19,21,24-26], where the elimination of hysteresis often diminishes the $-\Delta S_m$ to one-half or one-third of its giant value; outlining a formidable challenge in magnetic cooling technology.

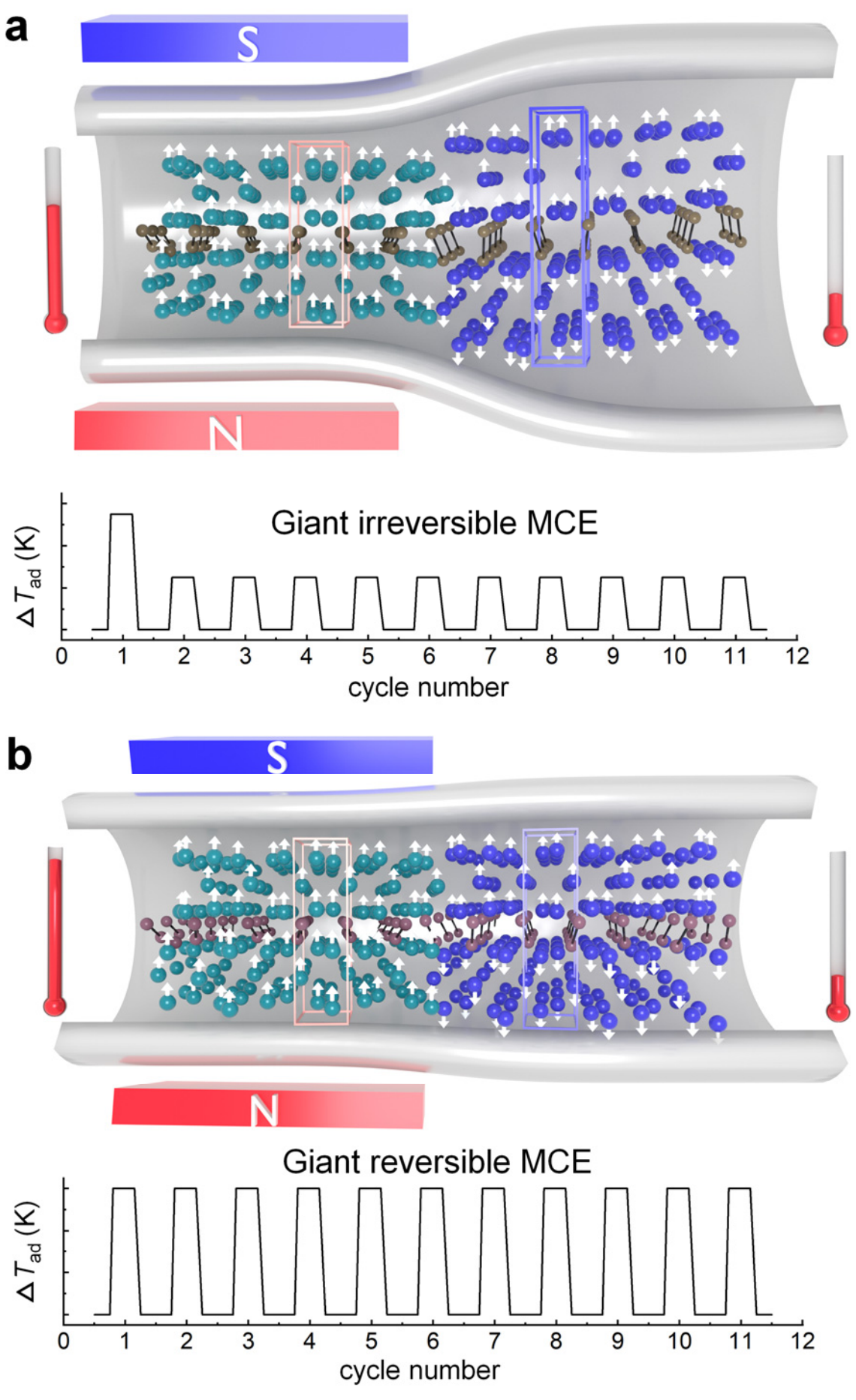

**Figure 1. Strategy for achieving giant and reversible MCE. a,** Limitation of first-order magnetic phase transition (FOMT): The $Gd_5Ge_4$ compound undergoes FOMT, transitioning from an antiferromagnetic to a ferromagnetic state, involving a transition between two orthorhombic structures with different lattice constants. This transition results in a giant magnetocaloric effect (MCE) in an adiabatic temperature change ($\Delta T_{ad}$) observed in thermometers. However, the giant MCE cannot be maintained in cyclic operation owing to large thermal hysteresis. **b,** Achieving giant and reversible MCE via covalent bond engineering: This approach involves compositional modifications in covalent bonds within a unit cell. For example, in the $Gd_5Sn_2Ge_2$ compound, the composition of covalent bonds changes from Ge–Ge to Sn(Ge)–Sn(Ge) bonds. The change in crystal structure upon transition is depicted using a unit cell with blue and red boxes representing two orthorhombic phases. Gd atoms in two orthorhombic structures are denoted by green and blue spheres, with white arrows indicating their magnetic moments. Ge and Sn(Ge) atoms are represented by yellow and pink spheres, respectively, connected by black covalent bonds. The $Gd_5Sn_2Ge_2$ compound is alloyed, with Sn atoms randomly distributed throughout the structure, including both inter-slab and intra-slab regions. Thus, the Sn(Ge)-Sn(Ge) bond description represents an average over Ge-Ge, Sn-Sn, and Ge-Sn bonds in the inter-slab region. For simplicity, only Ge3 and Sn(Ge)3 atoms are shown in the figure.

In many magnetocaloric materials, atomic-scale structural features, such as (covalent) bonds within the unit cell, play a critical role in shaping their magnetocaloric properties. This is particularly evident in systems like Ni-Mn-based Heusler alloys[27,28], $(Mn,Fe)_2(P,Si)$ [25,29-30] and $La(Fe,Si)_{13}$[31-32]compounds, where subtle interactions at the subunit-cell level have significant macroscopic effects. This necessitates a nuanced control at the subunit-cell-level to explore macroscopic properties. In the $Gd_5Ge_4$ compound, the magnetostructural phase transition is governed by the interplay between the atomic features of covalent bonds and magnetism. Specifically, the transition from antiferromagnetic to ferromagnetic states involves a change in the $Ge_3$–$Ge_3$ covalent bond length between two slabs, from 3.62 to 2.62 Å, across two respective orthorhombic structures, leading to hysteresis and irreversible functionality[33,34](**Figure 1a**). This unique layered slab structure within the unit cell provides a fertile platform to explore the possibility of eliminating hysteresis by engineering at finer structural scales—without compromising the GMCE. Herein, we demonstrate, as a proof of concept, that substituting Sn for Ge in the FOMT $Gd_5Ge_4$ compound enables the tuning covalency of bonds at the inter-slab region, modifying the phase transition to eliminate hysteresis while preserving latent heat for a giant MCE (**Figure 1b**). Our comprehensive experimental and theoretical investigation demonstrates that a fully reversible giant MCE can be realized through covalency tuning by chemical engineering at the sub-unit-cell scale, establishing a new design principle for effective control of hysteresis in magnetocaloric materials. These findings open transformative avenues for the rational design of next-generation magnetic refrigerants, particularly for practical applications in liquefaction of gases such as $H_2$, $N_2$, and natural gas, providing a green solution for efficient cooling technologies.

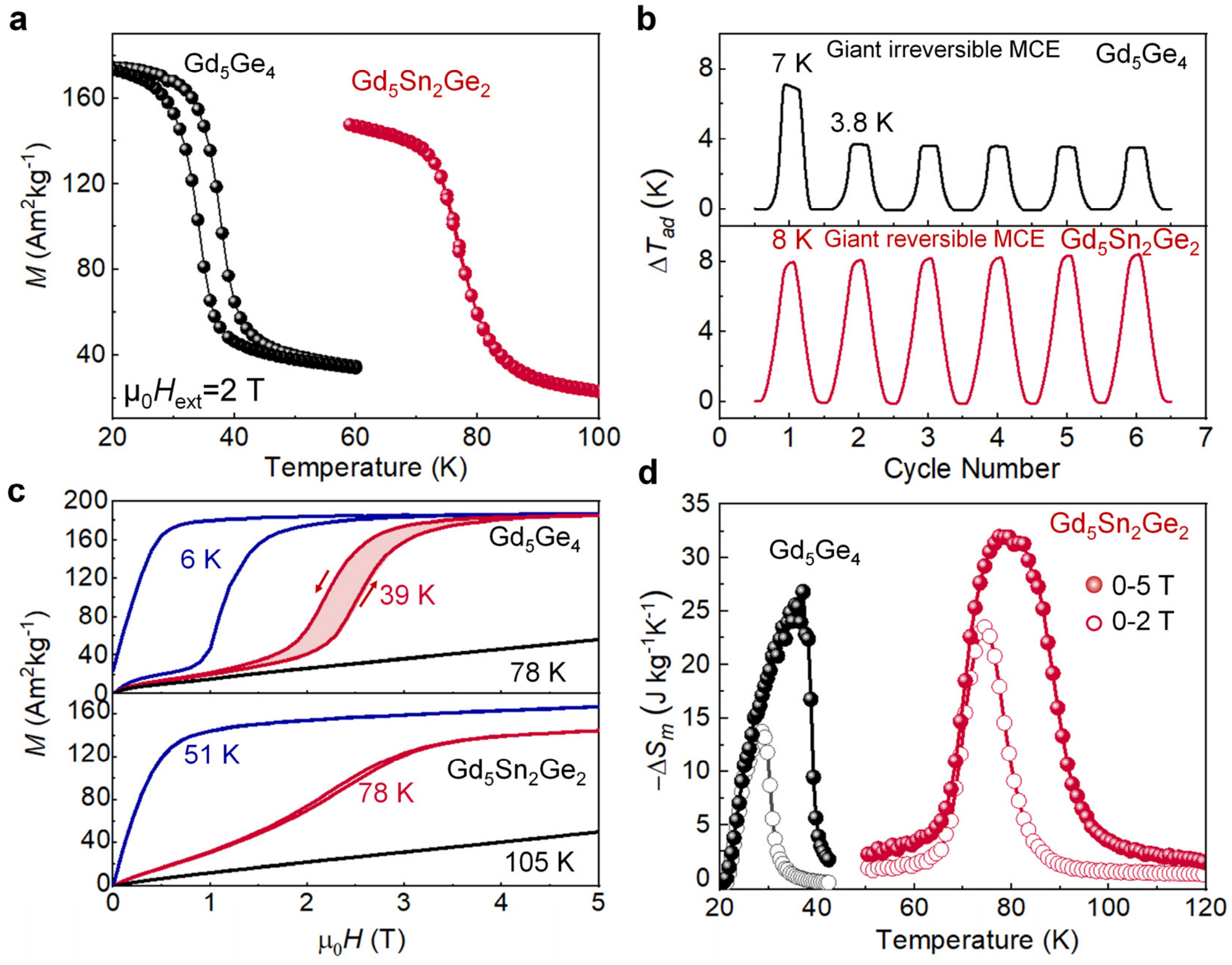

**Fig. 2: Advanced MCEs in the $Gd_5Sn_2Ge_2$ compound. a,** Magnetization (*M*) is plotted versus temperature at an applied magnetic field of 2 T. **b,** Adiabatic temperature change ($\Delta T_{ad}$) of $Gd_5Ge_4$ compound (upper panel) shows a giant MCE in $\Delta T_{ad}$, which cannot be maintained in cyclic fields. In contrast, we achieve a giant and reversible MCE in the $Gd_5Sn_2Ge_2$ compound (lower panel of **b**). **c,** Energy loss caused by hysteresis for $Gd_5Ge_4$ and $Gd_5Sn_2Ge_2$ compounds, which is numerically calculated using the enclosed area. **d,** $-\Delta S_m$ under a magnetic field change of 2 and 5 T measured for $Gd_5Ge_4$ and $Gd_5Sn_2Ge_2$ compounds using multiple *M-T* curves and the Maxwell equation. The adiabatic temperature change was directly measured under adiabatic conditions by a thermometer, which is confirmed via indirect measurement in **supplementary Fig. S1 (**Supporting Information**).**

## 2. Results and Discussion

### 2.1. Approaching hysteresis-free transition

The $Gd_5Ge_4$ compound undergoes a magnetostructural phase transition at 39 K under an external field of 2 T (**Figure 2a**), resulting in a GMCE in magnetic entropy change ($-\Delta S_m$)[33-34]. However, unlike $-\Delta S_m$, adiabatic temperature change ($\Delta T_{ad}$)—the equally important critical parameter for efficient MCE performance is underreported due to experimental challenges. In particular, direct measurements of $\Delta T_{ad}$ under cyclic conditions remain scarce, despite being essential for assessing the reversibility of MCE. In this work, we directly and cyclically measured $\Delta T_{ad}$ to rigorously assess MCE reversibility in $Gd_5Ge_4$ compound. Upon initial application of a 5 T magnetic field, the compound exhibits a $\Delta T_{ad}$ of 7 K associated with its first-order phase transition (**Figure 2b**). However, the large cooling effect is not sustained in the subsequent cycles because of thermal hysteresis of approximately 5 K (**Figure 2a**). Practically, only a $\Delta T_{ad}$ of 3.8 K can be maintained. This phenomenon poses a challenge for all magnetocaloric materials undergoing FOMT[7,35,36], thus hindering their practical applications. The large thermal hysteresis observed in the $Gd_5Ge_4$ compound arises from the interplay between covalent bonding at the sub-unit cell scale and magnetism[33,34]. To address this issue, sub-unit cell-level engineering is required, and chemical substitution presents a potential solution. In this study, we demonstrate this by substituting Ge with Sn in the $Gd_5Ge_4$ compound. As shown in **Figure 2a**, a hysteresis-free transition can be achieved in the $Gd_5Sn_2Ge_2$ compound; the MCE in $\Delta T_{ad}$ and $-\Delta S_m$ are simultaneously improved to 8 K and 32 $J \cdot kg^{-1} \cdot K^{-1}$, respectively. Notably, the GMCE in $\Delta T_{ad}$ robustly withstands cyclic operation via elimination of hysteresis (**Figure 2b**). It should be emphasized that the elimination of hysteresis in this work does not compromise the GMCE that is typically observed in FOMT materials, presenting a rare combination of benefits. The magnetic hysteresis is further quantified as the enclosed area between the ascending and descending magnetization curves (shaded area in **Figure 2c**), representing the energy lost as dissipated heat in a magnetic refrigeration cycle[24,37]. The developed $Gd_5Sn_2Ge_2$ compound exhibits a negligible hysteresis loss of 9 $J \cdot kg^{-1}$, which is only one-sixth of the loss observed for $Gd_5Ge_4$ compound (55 $J \cdot kg^{-1}$ at 39 K). This highlights the significant advantage of hysteresis elimination. While the end-member $Gd_5Sn_4$ compound[38] demonstrates a large $-\Delta S_m$, its $\Delta T_{ad}$ and cyclic stability remain unexplored, likely due to its air sensitivity. Similarly, Sn substitution in the $Gd_5(Ge,Si)_4$ system is known to enhance MCE in $-\Delta S_m$ and reduce hysteresis[39-41]; however, its reversibility under cyclic conditions and the underlying mechanisms remains poorly understood, partly due to the compositional complexity, which complicates theoretical modeling of local chemical effects on macroscopic behavior. Uncovering this hidden mechanism could unlock new strategies for designing materials that combine both giant and reversible MCEs. In contrast, the $Gd_5Sn_2Ge_2$ compound developed in this work provides a chemically well-defined platform, enabling accurate investigation of how sub-nanoscale chemical modifications influence reversibility of MCE.

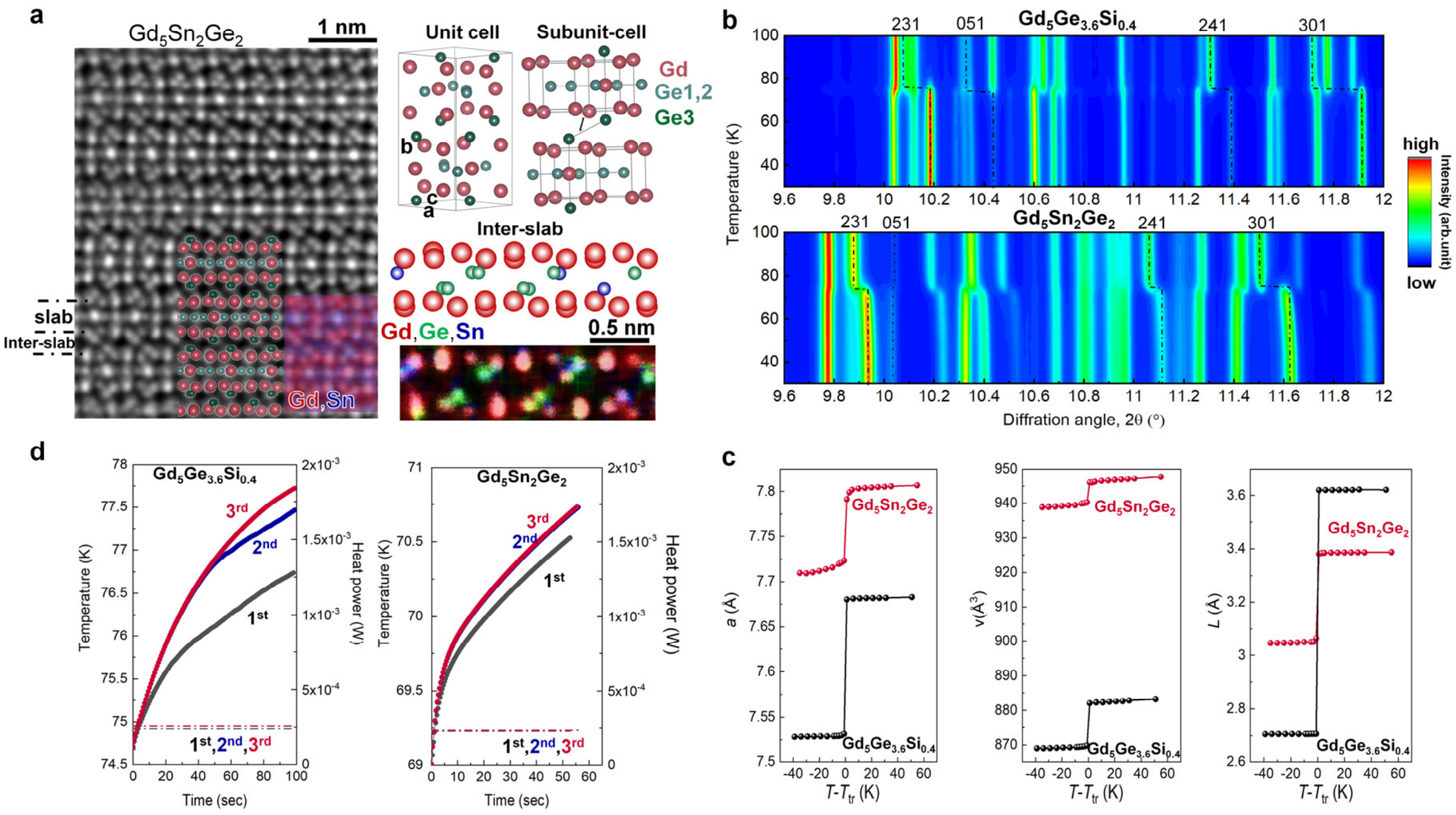

**Fig. 3: Characterization of atomic features at and within unit-cell scales and their response to external stimuli. a,** Crystal structure of a unit cell and bonds within a unit cell characterized via transmission electron microscopy. A unit cell of *pnma* crystal structure is depicted for $Gd_5Ge_4$ compound, with red spheres representing Gd atoms and green spheres representing Ge atoms. Ge occupies three nonequivalent sites: $Ge_1$ (light green spheres) and $Ge_2$ (light green spheres) are located within the slab, while $Ge_3$ (dark green spheres) is situated in the inter-slab region, with the $Ge_3$–$Ge_3$ bond connecting two slabs, bond length *l*. High-resolution high-angle annular dark field (HAADF) scanning transmission electron microscopy (STEM) image acquired along the [101] projection, with superimposed crystal structure (only showing Ge atom for simplicity) and energy-dispersive X-ray spectroscopy (EDS) maps showing the distributions of Gd and Sn in the $Gd_5Sn_2Ge_2$ compound, this indicates that Sn replaces Ge at both intra-slab ($Ge_1$, $Ge_2$) and inter-slab ($Ge_3$) regions. Atomic arrangement within the inter-slab region of the $Gd_5Sn_2Ge_2$ compound from crystal structure and determined from EDS analysis focused on inter-slab area for constituent elements, Gd (red), Sn (blue) and Ge (green). **b,** Temperature-dependent synchrotron X-ray diffraction (S-XRD) patterns are presented for $Gd_5Ge_{3.6}Si_{0.4}$ (upper panel) and $Gd_5Sn_2Ge_2$ (lower panel), with the black dashed line illustrating the characteristic diffraction plane shift during the phase transition. **c,** Temperature-dependence of the lattice parameters (*a*, volume (*v*) of a unit cell) and bond length *l* are obtained from the Rietveld refinement of S-XRD data. The *x*-axis represents the temperature minus transition temperature ($T_{tr}$). **d,** Time evolution of the sample temperature obtained using the relaxation method at the transition temperature for $Gd_5Ge_{3.6}Si_{0.4}$ and $Gd_5Sn_2Ge_2$ compounds. The dashed lines denote heat power for each measurement (1st (black), 2nd (blue) and 3rd (red) measurements).

## 2.2. Deciphering the phase transition

Here, we investigated the structural details at the unit-cell and subunit-cell scales and examined their responses to external stimuli. The $Gd_5Ge_4$-based compound, with an orthorhombic structure (*Pnma*) (**Figure 3a**), consists of a layered architecture (slab) interconnected by $Ge_3$–$Ge_3$ covalent bonds[22,34]. Real-space imaging and spectroscopy via transmission electron microscopy allow for the direct observation of these structures at the atomic scale. **Figure 3a** shows an element-resolved map revealing Sn substitution for Ge within both the intra-slab and inter-slab regions. In the inter-slab region, the Sn atoms are resolved and we use the notation $Sn(Ge)_3$–$Sn(Ge)_3$ to represent an averaged bond description, encompassing $Ge_3$–$Ge_3$, $Sn_3$–$Sn_3$, and $Ge_3$–$Sn_3$ bonds within the inter-slab region of the $Gd_5Sn_2Ge_2$ compound. This leads to changes in both the lattice constant and bond length between the two slabs in the orthorhombic crystal structure, as refined through synchrotron X-ray diffraction (S-XRD) patterns (**Supplementary note 1, Supporting Information**). Temperature-dependent S-XRD was then used to trace the thermally induced phase transitions in detail. It's noteworthy that the $Gd_5Ge_4$ compound exhibits a unique phenomenon, known as a kinetically arrested transformation (**Supplementary Fig. S2, Supporting Information**). This requires an additional external field of 2 T to induce structural change[34]. Instead, a trace substitution of Si for Ge does not alter the strong characteristics of the FOMT[22], as a similar thermal hysteresis of 5 K was recorded for $Gd_5Ge_{3.6}Si_{0.4}$ compound, but enables a field-free phase transition driven solely by temperature changes (**Supplementary Fig. S3, Supporting Information**). Such consistency allows for the exploration of field-free thermally driven phase transitions to understand the underlying mechanism of hysteresis elimination in the $Gd_5Sn_2Ge_2$ compound compared to the $Gd_5Ge_{3.6}Si_{0.4}$ compound.

**Figure 3b** compares the *in-situ* S-XRD measurements of $Gd_5Ge_{3.6}Si_{0.4}$ and $Gd_5Sn_2Ge_2$ compounds, showing distinct distortions near the transition temperatures owing to a structural transition between the two orthorhombic polymorphs in both samples. The shift in the characteristic diffraction plane (**Figure 3b**) indicates a change in the lattice spacing of the unit cell (**Figure 3c**). During the cooling process, the high-temperature orthorhombic (O(II)) phase transforms into the low-temperature orthorhombic (O(I)) phase with a shear displacement along the *a*-axis[34], resulting in approximately 2% shrinkage of the lattice constant, *a*, for the $Gd_5Ge_{3.6}Si_{0.4}$ compound. The change in *a* is more pronounced than those in the *b* and *c* lattice constants (**Supplementary Fig. S4, Supporting Information**). In contrast, the $Gd_5Sn_2Ge_2$ compound exhibits a smaller change in lattice spacing and unit cell volume during the transition. The shear displacement during the phase transition is further characterized by a change in the length of the Ge3–Ge3 covalent bond in the $Gd_5Ge_4$-based system. For the $Gd_5Ge_{3.6}Si_{0.4}$ compound, the bond length (*l*) changes from 3.6 Å for the O(II) phase to 2.7 Å for the O(I) phase, which is consistent with that observed for the $Gd_5Ge_4$ compound in the transition process driven by a magnetic field[34]. In contrast, the formation of Sn(Ge)$_3$–Sn(Ge)$_3$ bonds in the $Gd_5Sn_2Ge_2$ compound resulted in a small change of 0.3 Å in *l*. The presence of latent heat is another signature of first-order phase transition, evidenced by the time evolution of the sample temperature near the transition temperature (**Figure 3d**). For the two studied samples, the temperature rise during the initial run (1$^{st}$ run) was substantially suppressed due to the existence of latent heat, compared to subsequent runs (2$^{nd}$ and 3$^{rd}$ runs). However, the degree of suppression in the 1$^{st}$ run for $Gd_5Sn_2Ge_2$ compound was significantly lower than that observed for the $Gd_5Ge_{3.6}Si_{0.4}$ sample. This indicates that the chemical substitution of Sn for Ge can substantially weaken the nature of first-order phase transition. Consequently, our combined analysis of S-XRD and latent heat measurements confirmed that structural change and its associated latent heat were suppressed (but not completely eliminated) in the $Gd_5Sn_2Ge_2$ sample. This suppression was sufficient to eliminate hysteresis, resulting in a non-hysteretic transition while maintaining structural change for a GMCE. It should be noted further substitution of Si for Ge in the $Gd_5Sn_2Ge_{0.8}Si_{1.2}$ compound eliminates latent heat (**Supplementary Fig. S5, Supporting Information**), leading to a smaller $-\Delta S_m$ of 18 J·kg$^{-1}$·K$^{-1}$.

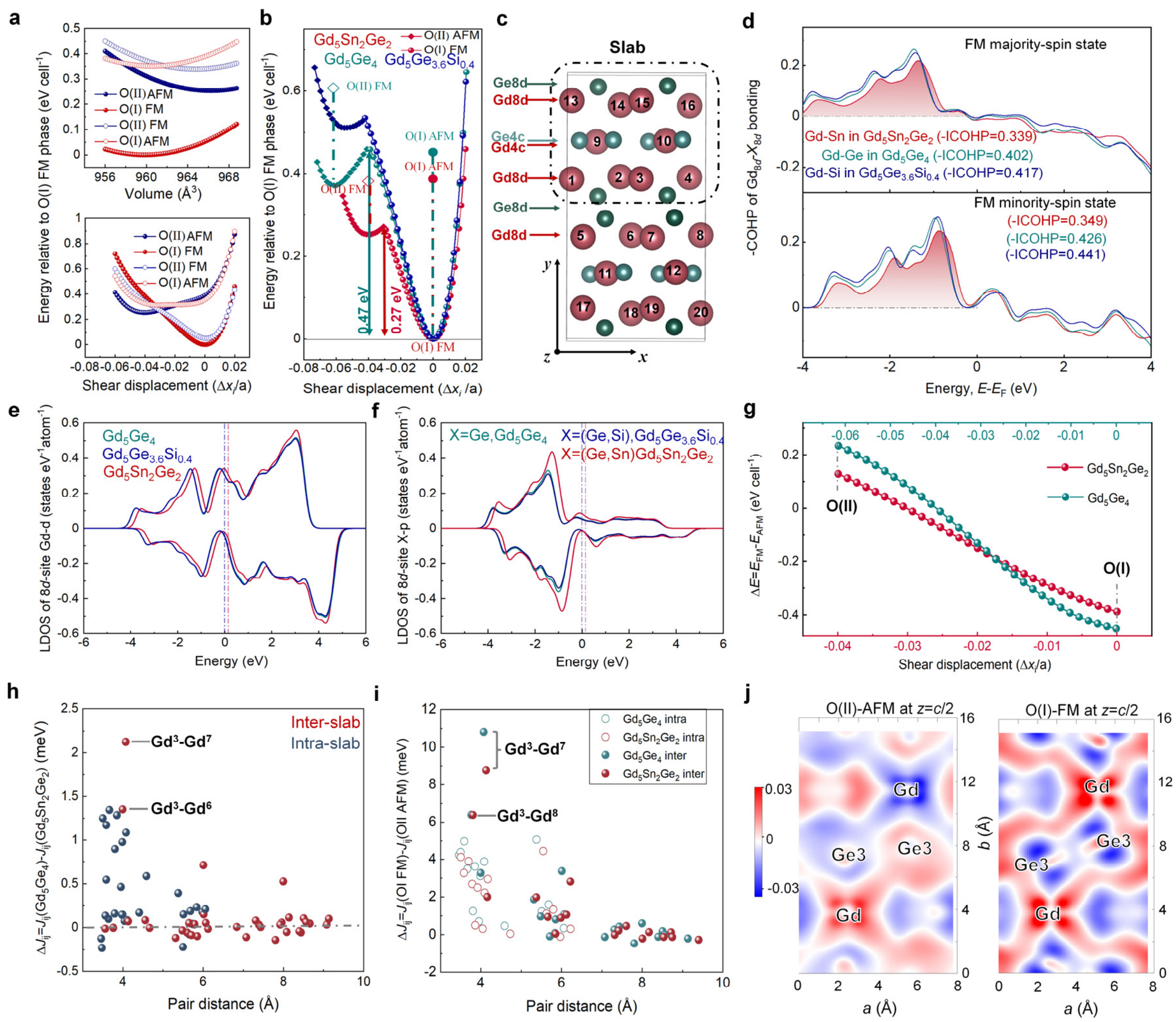

**Fig. 4: First principles calculations revealing the role of covalent bonds within a unit cell in shaping the magneto-structural phase transition and magnetism. a,** Relative energy as a function of volume, with fixed atomic positions for each structure and relative energy as a function of shear displacement of neighboring slabs, with the shear displacement along $a$-axis in fractional coordinates $\Delta x_i/a$, while the volumes are fixed for each structure. We consider the two structures, O(I) and O(II), each with two different magnetic configurations: ferromagnetic (FM) and antiferromagnetic (AFM). The relative energy is the energy with respect to energy minimum of the O(I) FM phase, In the lower panel, we set the minimum of the O(I) ferromagnetic structure as 0 for $x$-axis. **b,** Energy relative to the energy minimum of the O(I) FM phase as a function of shear displacement along $a$-axis in fractional coordinates $\Delta x_i/a$, for $Gd_5Ge_4$, $Gd_5Ge_{3.6}Si_{0.4}$ and $Gd_5Sn_2Ge_2$ compounds. We set the minimum of the O(I) ferromagnetic structure as 0 for $x$-axis in each studied compound. The activation barriers, $\triangle E$, from the O(I) FM to O(II) AFM phases are determined to be 0.27 and 0.47 eV for $Gd_5Sn_2Ge_2$ and $Gd_5Ge_4$ compounds, respectively, which are illustrated by solid red and green arrow. To examine the effect of exchange interaction, the positions of O(II) FM (open squares) and O(I) AFM (solid spheres) for $Gd_5Ge_4$ (green) and $Gd_5Sn_2Ge_2$ (red) compounds in the energy landscape are also illustrated, the dashed lines denote the exchange interaction for O(I) and O(II) structures in $Gd_5Ge_4$ (green) and $Gd_5Sn_2Ge_2$ (red) compounds. **c,** The unit cell of the *Pnma* crystal structure. Red spheres represent Gd atoms, while light and dark green spheres represent Ge atoms at the 4$c$-site and 8$d$-site, respectively. The $Ge_3$ sites, which

occupy the 8*d* position, are located in the inter-slab region, whereas $Ge_1$ and $Ge_2$ occupy the 4*c* sites in the intra-slab region. Gd atoms in the unit cell are labeled in the unit cell; for example, the eight Gd atoms at the inter-slab region are designated as $Gd^1$ through $Gd^8$, the dash black square shows a slab within unit cell. **d,** Energy-dependent crystal orbital Hamilton population (COHP) of Gd–X (where X = Ge, Si, Sn) at the 8*d*-site in the inter-slab region, the red shaded area represent integration of the -COHP (-ICOHP) for Gd-Sn in $Gd_5Sn_2Ge_2$ compound and their -ICOHP values are listed in **d**. **e,** Energy-dependent local density of states (LDOS) for the 8*d*-site Gd-*d* orbitals in $Gd_5Ge_4$ (green), $Gd_5Ge_{3.6}Si_{0.4}$ (blue), and $Gd_5Sn_2Ge_2$ (red) at O(I) FM state. **f,** Energy-dependent LDOS for the 8*d*-site Ge-*p* orbital in $Gd_5Ge_4$ (green), (Ge,Si)-*p* orbital in $Gd_5Ge_{3.6}Si_{0.4}$ (blue), and (Sn,Ge)-*p* orbital in $Gd_5Sn_2Ge_2$ (red) at O(I) FM state. Due to structural differences in these three compounds, the Gd s-core state is used as a reference to align the band positions, resulting in a 0.1368 eV shift to higher energy and 0.01652 eV shift to lower energy for $Gd_5Sn_2Ge_2$ and $Gd_5Ge_{3.6}Si_{0.4}$ compounds, respectively, compared to $Gd_5Ge_4$ compound. The dashed line indicates the Fermi level for the $Gd_5Ge_{3.6}Si_{0.4}$ (blue) and $Gd_5Sn_2Ge_2$ (red) compounds. (Sn,Ge) and (Ge,Si) denote the Sn/Ge and Ge/Si mixtures, respectively, based on their compositions. **g,** Exchange interaction energy ($\Delta E = E_{FM}$-$E_{AFM}$) between the ferromagnetic (FM) and antiferromagnetic (AFM) states across calculation steps transitioning from the O(II) structure to the O(I) structure, with varying shear displacement, $\Delta x_i/a$. Since the shear displacement for two compounds is different, the top *x*-axis and bottom *x*-axis corresponds to $Gd_5Ge_4$ and $Gd_5Sn_2Ge_2$ compounds, respectively. We set the O(I) FM structure as 0 for *x* axis. **h,** Pairwise comparison of exchange interactions in $Gd_5Ge_4$ and $Gd_5Sn_2Ge_2$ compounds, $\Delta J_{ij}=J_{ij}(Gd_5Ge_4)$-$J_{ij}(Gd_5Sn_2Ge_2)$, the difference of pairwise $J_{ij}$ between $Gd_5Ge_4$ and $Gd_5Sn_2Ge_2$ compounds. "$Gd^3$-$Gd^7$" refers to the pair interaction between atoms $Gd^3$ and $Gd^7$ as shown in **c**. Equivalent pairs, such as pairs $Gd^1$-$Gd^5$, $Gd^{13}$-$Gd^{17}$, and $Gd^{15}$-$Gd^{19}$, share the same pair distance and exchange interaction constant ($J_{ij}$) with $Gd^3$-$Gd^7$ due to periodicity and local symmetry (**Supplementary Table 3, Supporting Information**). The *x*-axis indicates the pair distance in the $Gd_5Ge_4$ compound. **i,** Pairwise comparison of difference in exchange interactions between the O(I) FM and O(II) AFM states in the $Gd_5Ge_4$ and $Gd_5Sn_2Ge_2$ compounds, with the *x*-axis showing the pair distance in FM state of $Gd_5Ge_4$ and $Gd_5Sn_2Ge_2$ compounds. The open circles stand for difference of intra-slab interaction and solid circles represent difference of inter-slab interaction. **j,** Spin density for the O(II) AFM and O(I) FM phases for the $Gd_5Sn_2Ge_2$ compound is depicted with a cut at $z = c/2$ of the crystallographic structure. The positions of intra-slab Gd and inter-slab $Sn(Ge)_3$ are labeled as Gd and $Ge_3$, respectively, in the spin density map. In the O(II) AFM state, two intra-slab Gd exhibit antiferromagnetic coupling, whereas in the O(I) FM state, they show ferromagnetic coupling. The two $Sn(Ge)_3$ atoms are isolated at the O(II) AFM state, indicative of broken bonds, while they are connected in the O(I) FM state.

### 2.3. Mechanism of thermo-magnetic phase transitions

To investigate the role of structural intricacies in controlling magnetism and magnetostructural phase transitions, we conducted first-principles calculations. In the context of Landau theory[42,43], the FOMT theoretically hinges on the energy barriers that separate the local and global minima in the Gibbs free energy. For $Gd_5Sn_2Ge_2$ compound, we resolved the phase transition between the two polymorphs, resulting in changes in the unit-cell volume and covalent bond lengths within the unit cell (**Figure 3c**). Here, we examined the influence of two potential factors—volume and shear displacement between $Sn(Ge)_3$–$Sn(Ge)_3$ bonds—on inducing the phase transition with different magnetic configurations. The observed lowest energies of the O(I) ferromagnetic (FM) phase at every volume (**Figure 4a**) indicate that the phase transition cannot be attributed to volume changes alone. In contrast, the shear displacement of the Sn(Ge)3–Sn(Ge)3 bond could trigger the phase transition (**Figure 4a**), as evidenced by two local energy minima for the O(II) antiferromagnetic (AFM) and O(I) FM phases observed at approximately $\Delta x_i/a$ = 0 and -0.04 (approximately shear displacement of -0.31 Å along *a* axis with respect to the minima of O(I) FM phase), respectively. The corresponding $Sn(Ge)_3$–$Sn(Ge)_3$ bond lengths are determined to be 3.59 and 2.96 Å for O(II) AFM and O(I) FM phases, respectively, which agrees with our experimental results (**Supplementary Table 1 and 2, Supporting Information**). To further understand the influence of chemical composition on the magnetostructural phase transitions, we studied this system for different compositions. In all studied compounds, the phase transition from O(I) FM to O(II) AFM is observed and associated with the shear displacement of covalent bonds (**Supplementary Fig. S6, Supporting Information**); the change in bond length during the phase transition is smaller for the $Gd_5Sn_2Ge_2$ compound (**Supplementary Fig. S7, Supporting Information**). This calculation replicated experimental observations. Interestingly, the activation barrier ($\triangle E$) from the O(I) FM phase to the O(II) AFM phase reduce from 0.47 for $Gd_5Ge_4$ compound to 0.27 $eV \cdot cell^{-1}$ for the $Gd_5Sn_2Ge_2$ compound (**Figure 4b, Supporting Information**), indicating an energetically more commensurate phase transition for $Gd_5Sn_2Ge_2$ compound. This elucidates the elimination of hysteresis presented in this work. Further investigation reveals that the Sn concentration in the $Sn(Ge)_3$–$Sn(Ge)_3$ bonds within the inter-slab region significantly influences both the activation energy and the magnetostructural phase transition behavior (**Supplementary note 2, Supporting Information**).

To elucidate how Sn substitution in $Gd_5Ge_4$ reduces activation energy, we analyzed the bonding nature using crystal orbital Hamilton population (COHP) calculations. Experimentally, changes in the bond length at the inter-slab region were expected to influence the 8d-site Gd-$X_3$ (X = Ge, Si, Sn) bonds and their covalency due to hybridization between the *d* orbitals of Gd and *p* orbitals of $X_3$. COHP calculations of the Gd-$X_3$ bonds are shown in **Figure 4d**. The bond strength (covalency) can be evaluated by the integration of the COHP with an energy range of -4.0 eV to Fermi energy. Integration of the -COHP (-ICOHP) for the O(I) FM structure reveals that the Gd-Sn bond has a smaller -ICOHP of 0.339 (0.349) for majority (minority)-spin states, compared to 0.402 (0.426) for the Gd-Ge bond and 0.417 (0.441) for the Gd-Si bond.

Similar results are obtained for the O(II) AFM structure (**Supplementary Fig. S8, Supporting Information**). These results suggest that Sn doping reduces covalency of Gd-X at inter-slab region in $Gd_5Ge_4$ compound, possibly due to the spatial broadening of the Sn-5*p* wavefunction relative to the Ge-4*p* wavefunction. Structural changes are also anticipated to affect the local density of states (LDOS) of the studied compounds. As shown in **Figures 4e and 4f**, the LDOS of the 8d-site Gd-*d* and Ge(Sn,Si)-*p* states in the O(I) FM phase of $Gd_5Ge_2Sn_2$ shifts to higher energies compared to those in $Gd_5Ge_4$ and $Gd_5Ge_{3.6}Si_{0.4}$ compounds. The bonding state at around energy level of -1.3 eV shift to higher energy also confirms the weaker covalency in $Gd_5Sn_2Ge_2$ compound. Similar energy shifts are also observed in the AFM state (**Supplementary Fig. S9, Supporting Information**). These changes in covalency are expected to further influence exchange interactions. We then calculated the energy difference between the FM and AFM states ($\Delta E = E_{FM} - E_{AFM}$) to quantify the exchange interaction. For $Gd_5Ge_4$ compound, the exchange interaction energies for the O(I) and O(II) structures are -0.45 and 0.24 $eV\cdot cell^{-1}$, respectively (**Figure 4g**). The exchange energy is further illustrated by positioning the energies of O(I) AFM and O(II) FM within the energy landscape in **Figure 4b**, which shows that the energy profile can be strongly influenced by exchange interactions. With Sn substitution, the magnitude of exchange interaction energy decreases by 0.07 $eV\cdot cell^{-1}$, from -0.45 to -0.38 $eV\cdot cell^{-1}$ for the O(I) structure, and 0.11 $eV\cdot cell^{-1}$, from 0.24 to 0.13 $eV\cdot cell^{-1}$ for the O(II) structure (**Figure 4g**). This exchange interaction reduction can explain the overall reduction in activation energy, which decreases from 0.47 to 0.27 $eV\cdot cell^{-1}$ due to Sn substitution (**Figure 4b**).

We further examined exchange interaction constants ($J_{ij}$) within (intra-slab) and between neighboring slabs (inter-slab) for the 20 Gd atoms in the unit cell (**Figure 4c**) using SPR-KKR code, with pairwise $J_{ij}$ data detailed in **Supplementary Table 3** (Supporting Information). **Figure 4h** compares the difference of pairwise exchange interactions between $Gd_5Ge_4$ and $Gd_5Sn_2Ge_2$ compounds, indicating that interactions are generally stronger in $Gd_5Ge_4$ compound, agreeing with calculations using density functional theory (**Figure 4g**). The Sn substitution significantly impacts $Gd^3$-$Gd^7$ and $Gd^3$-$Gd^6$ pairs, connecting Gd atoms across slabs. This highlights the importance of chemical engineering in the inter-slab region. The exchange interaction difference estimated using $J_{ij}$, between the relaxed O(I) FM phase and O(II) AFM phase shows a reduction from 0.374 $eV\cdot cell^{-1}$ in the $Gd_5Ge_4$ compound to 0.291 eV $cell^{-1}$ in the $Gd_5Sn_2Ge_2$ compound, independently verifying that Sn alloying weakens exchange interactions and reduces activation energy. We then explored the how the inter-slab exchange interactions influence the AFM-to-FM transition. As depicted in **Figure 4i**, the inter-slab $Gd^3$- $Gd^7$ and $Gd^3$-$Gd^8$ pairs exhibit largest change in $J_{ij}$ between O(II) AFM and O(I) FM states. During the phase transition process, the thermally driven displacement of inter-slab Ge(Sn) atoms modulates inter-slab interactions as indicated by the spin density contour map (**Figure 4j**). This modulation drives the magnetism transition from AFM to FM state as shown by theoretical density of states (**Supplementary Fig. S10, Supporting Information**) and experimentally measured Gd moment using XMCD (**Supplementary note 3, Supporting Information**). These theoretical insights highlight the critical role of inter-slab bonds in controlling magnetostructural and magnetism transitions.

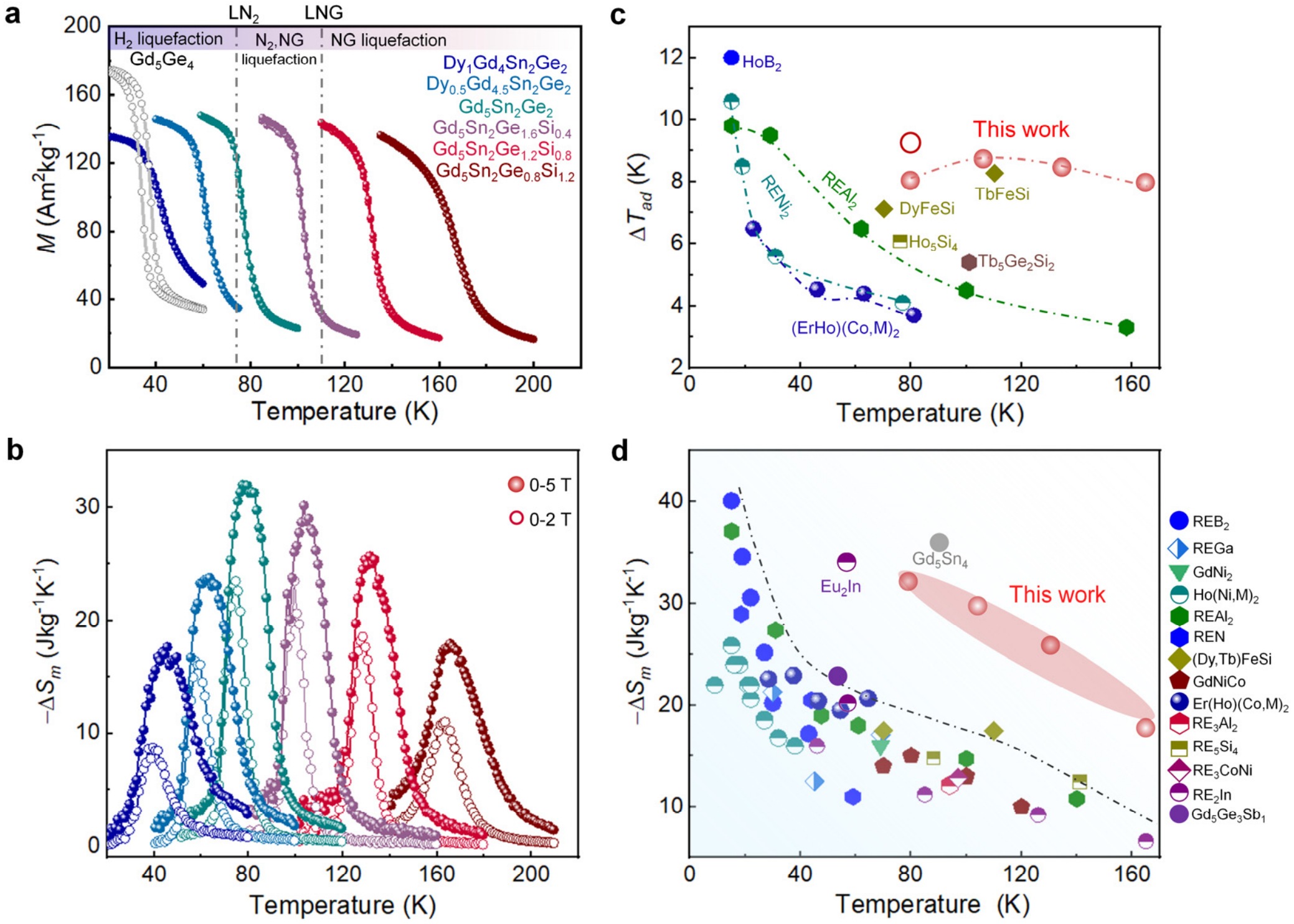

**Fig. 5: Versatile application for gas liquefaction by tailoring transition and comparing MCEs with those of widely researched magnetocaloric materials. a,** Temperature-dependent magnetisation $M(T)$ measurements of the cooling and heating branches at 1 K intervals ($\mu_0 H$ = 2 T) for a series of magnetic refrigerants with tailored transition temperatures, covering the temperature window required for application of $H_2$, $N_2$, and natural gas (NG) liquefaction. **b,** Temperature-dependent magnetic entropy change $-\Delta S_m(T)$ curves of magnetic refrigerants under a field change of 2 T (open circle) and 5 T (solid circle). The colors of the curves in **a**, **b** correspond to the color of the listed alloy compositions in **a**. **c,** Comparison of reversible adiabatic temperature change ($\Delta T_{ad}$) at the field change of 5 T is made with other magnetocaloric materials with transition temperature from 15 to 170 K. Indirect measurement of $\Delta T_{ad}$ for $HoB_2$ (ref. [44]), $Ho_5Si_4$ (ref. [45]), $Tb_5Ge_2Si_2$ (ref. [46]), $Gd_5Sn_2Ge_2$ compound (open red circle) and direct measurement for rare earth (RE)$Al_2$, $RENi_2$ (ref. [47,48]), $(Er,Ho)(Co,M)_2$ (ref. [21]) and $Gd_5Sn_2(GeSi)_2$ compounds (3D red spheres) of this study. **d**, Comparison of magnetic entropy change under a field change of 5 T for the developed compounds with other magnetocaloric materials with non-hysteretic transitions ranging from 15 to 170 K. The $-\Delta S_m$ for $Gd_5Sn_4$ (refs. [38]), $REB_2$ (refs. [44,49]), REGa

(refs. [50,51]), $GdNi_2$ (ref. [47]), $Ho(Ni,Co)_2$ (ref. [52]), $REAl_2$ (ref. [47]), REN (ref. [53]), (Dy,Tb)FeSi (ref. [54]), GdNiCo (ref. [55]), $(Er,Ho)(Co,M)_2$ (ref. [21]), $RE_3Al_2$ (ref. [56]), $RE_5Si_4$ (Refs. [45,46]), $RE_3CoNi$ (ref. [57]), $RE_2In$ (refs. [58–61]), $Gd_5Ge_3Sb_1$ (ref. [62]), $Eu_2In$ (ref. [63]) and $Gd_5Sn_2(GeSi)_2$ compounds of this study.

**2.4. Applicability of developed materials**

Another challenge in implementing magnetic cooling systems is the limited operating temperature range of the refrigerants, primarily because of the narrow temperature span of the GMCE. A solution to this constraint—based on active regeneration schemes—necessitates the tunability of the transition temperature in magnetocaloric materials. A broad transition temperature range of 40–160 K can be achieved by strategically substituting Dy for Gd and Si for Ge (**Figure 5a**). Furthermore, $-\Delta S_m$ above 18 $J{\cdot}kg^{-1}{\cdot}K^{-1}$ (**Figure 5b**) is maintained for these non-hysteretic transitions and $\Delta T_{ad}$ for all developed compounds remains stable under cyclic operation (**Supplementary Fig. S11, Supporting Information**) Next, we evaluated the superiorities of magnetocaloric performance across two crucial dimensions ($\Delta T_{ad}$, $-\Delta S_m$) as preferred by application and compared them with those of widely investigated refrigerants. The compounds developed herein exhibited an excellent combination of these favorable properties. **Figure 5c and d** show the reversible MCEs at 15–160 K for a wide range of materials[21,38, 44-63]. The absence of materials with a large MCE for both $\Delta T_{ad}$ and $-\Delta S_m$ above 50 K represents a material bottleneck for cryogenic applications in magnetic refrigeration technology. This challenge can be addressed by implementing the developed materials into a magnetic cooling system for gas liquefaction, as they exhibit giant MCEs in both $-\Delta S_m$ and reversible $\Delta T_{ad}$. Among these, the $Gd_5Sn_2Ge_2$-based compounds exhibited 1.5–2 times larger MCEs than most existing magnetic refrigerants in both $\Delta T_{ad}$ and $-\Delta S_m$. Although the $Gd_5Sn_4$ compound shows a large MCE around 90 K, its hysteresis behavior and associated reversibility under cyclic conditions have not yet been reported[38]. Similarly, $Eu_2In$ shows a larger $-\Delta S_m$ at 59 K, but the air-sensitive nature of both $Gd_5Sn_4$ and $Eu_2In$ compounds[38,63]undermines their applicability. In contrast, the developed $Gd_5Sn_2Ge_2$ compound demonstrates good phase stability under air exposure (**Supplementary Fig. S12, Supporting Information**). To further evaluate the properties, we consider another magnetocaloric metric, refrigeration capacity (RC) shown in **Supplementary note 4** (Supporting Information). A $-\Delta S_m$-RC diagram encompasses a diverse family of materials with transition temperatures spanning from 15 K to 320 K. The effective RCs for FOMT materials are confined to values below 400 $Jkg^{-1}$, falling short of the achievements of the materials developed here. Conversely, SOMT materials have higher RC values, but their magnetocaloric performance is overshadowed by their lower values of $-\Delta S_m$. The $Gd_5Sn_2Ge_2$ compounds in our work exhibit a unique combination of large effective RC of 550 Jkg$^{-1}$ and -$\Delta S_m$ of 32 Jkg$^{-1}$K$^{-1}$. In addition, the developed $Gd_5Sn_2Ge_2$ compound exhibit excellent mechanical stability during the cyclic operation and attractive cost performance (**Supplementary note 5, Supporting Information**). These comparisons underscore the potential of the developed compounds for the magnetic liquefaction of a range of gases, such as $H_2$, $N_2$, and natural gas.

### 3. Summary and outlook

In conclusion, we propose and demonstrate a novel approach to mitigate hysteresis in FOMT materials without sacrificing their magnetocaloric effect. The conventional approach is to achieve a crystallographically compatible phase transition by manipulating the overall unit cell properties. However, this method often faces a persistent trade-off between a giant MCE and good reversibility. Our approach involves precise control at the subunit-cell scale by the formation of $Sn(Ge)_3$–$Sn(Ge)_3$ bonds in the $Gd_5Ge_4$-based compound. This strategy promotes a favorable energy landscape for the magnetostructural phase transition by weakening of exchange interaction, unlocking the potential of FOMT magnetocaloric materials without introducing hysteresis-associated side effects. This delivers a synergistic achievement of magnetocaloric metrics. Furthermore, this approach could have broader implications for other magnetocaloric systems, where sub-unit-cell interactions critically influence magnetostructural behavior. Examples include NiMnX-based Heusler compounds[27,28], $(Mn,Fe)_2(Si,P)$ systems[25,29-30], and $La(FeSi)_{13}$ alloys[31,32]. In these materials, intricate interactions—such as p-d covalent bonding in Heusler alloys[27,28], Fe-Si covalent bonding in $Fe_2P$-type materials[25,29-30], and intra- and inter-cluster Fe-Fe bonding in $La(FeSi)_{13}$[31,32]—play critical roles in shaping magnetostructural phase transition. Moreover, the versatility of applications for the developed compounds has been demonstrated by their tunable, non-hysteretic phase transitions. A comprehensive evaluation of primary magnetocaloric properties establishes that the materials developed in this study hold significant promise for applications across a spectrum of magnetic refrigeration materials, particularly in the search of green alternative for gas liquefaction.

### 4. Methods

#### 4.1 Materials preparation and thermomagnetic measurements

Polycrystalline $Gd_5Ge_4$-based samples were synthesized via the arc melting of pure constituent elements in an Ar atmosphere, incorporating 1–5 wt.% additional rare earth to compensate for evaporation losses during sample preparation. The ingots were remelted four times after flipping to ensure homogeneity. A thermomagnetic analysis was performed using a superconducting quantum interference device magnetometer (SQUID-VSM). Temperature-dependent magnetization ($M$(T)) measurements were conducted on both the cooling and heating branches at 1-K intervals. The temperature-dependent magnetic entropy change ($-\Delta S_m$(T)) was calculated based on the following Maxwell's equation from multiple $M$-T curves at various external fields:

$$\Delta S_m = \mu_0 \int_H^0 \left(\frac{\partial M}{\partial T}\right)_H dH.$$

**4.2. Direct measurement of adiabatic temperature and latent heat**

To measure the adiabatic temperature change directly, a thermometer made of zirconium oxynitride thin-film CernoxTM (CX-SD, Lake Shore Cryotronics) was placed on the large surface of an approximately cubic sample and fixed by thin copper plates and thin copper wires. The background adiabatic temperature change contributed from the sample holder was subtracted considering its heat capacity. We inserted the sample assembly into a quantum design physical property measurement system (PPMS) manufactured by the Quantum Design Company to control the temperature and magnetic field. The sample space was pumped to obtain adiabatic conditions using a cryopump, and the pressure was maintained below $10^{-4}$ Torr. For latent heat measurements, we employed the relaxation method implemented in the PPMS. We evaluated the latent heat by measuring the time evolution of the sample temperature during measurement using the relaxation method. Details of this method are provided in **supplementary note 6** (Supporting Information).

**4.3. Indirect measurement of adiabatic temperature**

This property was stipulated after performing heat capacity measurements in the Quantum design PPMS. Heat capacity was measured in a plate-shaped sample using the 2τ method in applied magnetic fields of 0 T and 5 T.  The following equation allows to calculate entropy as a function of temperature in different magnetic fields:

$$S(T,H) = \int_0^T \left(\frac{C_P}{T}\right)_H dT$$

The indirect $\Delta T_{ad}$ is obtained considering an isoentropic difference of $S(T,H_0)$ and $S(T,H_f)$, where $H_0$ denotes the entropy curve for 0 T applied, and $H_f$ refers to the entropy with an applied field different than 0 T, as given in the following equation:

$$\Delta T_{ad} = T(S, H_f) - T(S, H_0)$$

**4.4 Structure characterizations**

The phase transition process was tracked using an *in-situ* synchrotron X-ray diffractometer (S-XRD) at the BL02B2 beamline of SPring-8. The measurement samples were ground into fine powders, placed in a borosilicate glass capillary (diameter, 0.3 mm and sealed with He gas. The temperatures of the powder samples were controlled using He and $N_2$ gas blowers. Temperature-dependent S-XRD patterns were collected during heating from 30 to 200 K. The samples were rotated during the measurements to improve the particle statistics. Six one-dimensional solid-state (MYTHEN) detectors were used to collect the XRD patterns. Diffraction data from 2° to ~ 78° were acquired via XRD measurements using multiple detectors at two designated detector angle positions (double-step measurements). The beam size was collimated to 0.5 mm (height) × 1.5 mm (width). The wavelength was determined to be 0.4956635 Å after conducting the calibration using $CeO_2$ standard sample. Scanning transmission electron microscopy (Spectra Ultra STEM, Thermo Fisher Scientific) equipped with energy-dispersive X-ray spectroscopy using four detectors was employed to observe the atomic-scale features. Drift correction was applied by the system during elemental mapping. The TEM samples were prepared using an FEI Helios G4-UX dual-beam system with the lift-out technique.

**4.5. Evolution of element-specific moment**

Soft X-ray magnetic circular dichroism (XMCD) measurements were performed using the BL25SU beamline at the SPring-8 synchrotron radiation facility. The XMCD spectra at the Gd $M_{4,5}$ edges were recorded at different temperatures during heating from 20 K in the ferromagnetic state to 160 K. A degree of circular polarization of 0.96 at 400 eV has been previously estimated[64] and is expected to be similar in the energy region used in this work. The spectra were recorded using the total electron yield method, with an angle between the incident X-ray beam and the magnetic field of 10 degrees[65]. Samples were prepared in a rod shape with a 1 $mm^2$ square section and 10 mm length and fractured in the ultra-high vacuum chamber of the XMCD system with a vacuum level of $< 5 \times 10^{-7}$Pa, allowing the measurement of the fresh surface. The XMCD signal ( $\mu_{XMCD}$ ) was obtained as $\mu_{XMCD} = (\mu_{l-} + \mu_{r+}) - (\mu_{l+} + \mu_{r-})$ where $\mu_l$ and $\mu_r$ represent the X-ray absorption spectrum (XAS) for the "left-handed" ($h_-$) and "right-handed" ($h_+$) helicity, respectively, and $\mu_+$ and $\mu_-$ represent the XAS for the positive and negative external applied magnetic field of 1.9 T. Magneto-optical sum rule analysis for XMCD[66–68]was used to calculate the magnetic moments of Gd, considering the spin correction factor for rare earths[69].

### 4.6. First-principles calculations

Spin-polarized density functional theory (DFT) calculations were performed using the projected augmented wave pseudopotential method as implemented in the Vienna *Ab-initio* Simulation Package code[70]. The exchange-correlation interaction was approximated using the Perdew–Burke–Ernzerhof[71] formulation based on the generalized gradient approximation (GGA). K-point sampling of the Brillouin zone was performed using a 6× 3 × 6 k-mesh grid. The spin-orbit coupling (SOC) was not consistently included in the calculations. The localized 4f electrons of Gd were treated based on the implementation of Hubbard's U parameter using the DFT + U method with U = 6.7 eV and J = 0.7 eV. The lattice constants and atomic positions of the O(I) and O(II) structures with FM and AFM states were determined by atomic relaxation in DFT calculations, and Sn and Si doping in $Gd_5Ge_4$ was simulated using a virtual crystal approximation[72]. A correction of the Columb interaction was made by giving an on-site U and J to the Gd f-orbital to obtain the correct Gd spin moment (7 μB). This has been done in previous studies[73]and considered to be one of the methods to treat the Gd f-orbital correctly in the DFT calculations. The values of U and J that we used were taken directly from the previous studies. Crystal Orbital Hamilton Population (COHP) analysis is conducted using the LOBSTER code[74].

Exchange interaction constants were computed using spin-polarized relativistic Korringa–Kohn–Rostoker (SPR-KKR) package[75,76], with optimized structures obtained by VASP. The k mesh for self-consistent and $J_{ij}$ calculations was chosen as 5 × 2 × 5 and 12 × 6 × 12, respectively.  The exchange-correlation interaction was approximated using the Perdew–Burke–Ernzerhof formulation[71]. Full-potential spin-polarized scalar-relativistic Korringa–Kohn–Rostoker method (FP-SP-SREL-KKR)[77]was employed that the spin-orbit coupling (SOC) was not included in the calculation.

**Supporting Information**

Supporting Information is available from the Wiley Online Library or from the author.

**Acknowledgments**

This study was in-part supported by the JSPS International Joint Research Program (JRP-LEAD with DFG; program number JPJSJRP20221608), JST ERATO "Magnetic Thermal Management Materials" (Grant No. JPMJER2201) and JSPS KAKENHI Grant Number JP19H05819 (to Y. Ma). Synchrotron radiation experiments were performed at BL02B2 and BL25SU of SPring-8 with the approval of the Japan Synchrotron Radiation Research Institute (JASRI) (Proposal Nos. 2023A1183, 2023A1510, and 2023A1730). H. Sepehri-Amin and T. Nakamura acknowledge the support from the NIMS-TOHOKU Joint Research Partnership Program. O. Gutfleisch and K. Skokov acknowledge financial support by the Deutsche Forschungsgemeinschaft (DFG) within the CRC/TRR 270 (Project-ID 405553726). A part of this work was supported by the Electron Microscopy Unit, National Institute for Materials Science (NIMS).

**Author Declarations**

The authors declare no competing financial interests.

**Author Contributions**

X. T. developed this idea and prepared the alloys. X. T. performed the magnetothermal measurements. N. T. performed adiabatic temperature changes and latent heat measurements. Y. Mi. conducted density functional theory (DFT) calculations and crystal orbital Hamilton population (COHP) analysis. E. X. and T. T. conduct the pairwise exchange interaction calculation using SPR-KKR code. S. K. and X. T. carried out *in-situ* synchrotron X-ray diffraction measurements and analyzed the data. T. O., X. T., A. M., and T. N. conducted soft X-ray magnetic circular dichroism (XMCD) measurements and analyzed the obtained XMCD spectra. X. T. conducted high-resolution transmission electron microscopy observations. D. A., K. S., and O. G. performed the indirect measurements of the adiabatic temperature change. Y. Ma. investigated the temperature dependence of the X-ray diffraction measurements. X. T. and H. S. interpreted key findings and wrote the manuscript. O. G., T. O., K. H. and H. S. supervised the study. All the authors discussed the results and commented on the manuscript.

**Data Availability Statement**

The data that support the findings of this work are available from corresponding author upon reasonable request.